\documentstyle[amssymb]{amsart}

\def\beq{\begin{equation}}
\def\eeq{\end{equation}}

\def\bq{\begin{quote}}
\def\eq{\end{quote}}

\newcommand{\be}{\begin{equation}}
\newcommand{\ee}{\end{equation}}

\newcommand{\al}{\alpha}
\newcommand{\pa}{\partial}
\newcommand{\ep}{\epsilon}

\newcommand{\la}{\lambda}

\newcommand{\om}{\omega}
\newcommand{\de}{\delta}

\newcommand{\De}{\Delta}

\begin{document}

\begin{flushright}
M\'exico ICN-UNAM 97-12
\end{flushright}
\vskip 1cm

\begin{center}
{\bf\Large\sf Two-body Elliptic Model in proper variables: Lie-algebraic
forms and their discretizations.}
\footnote{Contribution to the Proceedings of the Workshop on
Calogero-Moser-Sutherland models, Montreal, March 10-15, 1997}

\vskip 1cm
{\Large Alexander Turbiner{\normalsize
\footnote{e-mail: turbiner@@xochitl.nuclecu.unam.mx}${}^{,}
$\footnote{On leave of absence from the Institute for Theoretical
and Experimental Physics,  Moscow 117259,
Russia.}}}\\
{\em Instituto de Ciencias Nucleares, UNAM,
Apartado Postal 70-543,\\ 04510 Mexico D.F., Mexico}

\vskip 1cm

{\Large Abstract}
\end{center}

\begin{quote}
Two Lie algebraic forms of the 2-body Elliptic Calogero model are presented.
Translation-invariant and dilatation-invariant discretizations of the model
are obtained.
\end{quote}

\setcounter{equation}{0}
\section{Introduction}

The general $N$-body Lame operator is given by
\begin{equation}
L\equiv \De - n(n+1)\sum_{i>j}^N {\cal P}(x_i-x_j)  \ .
\end{equation}
where $n$ is positive integer and the Weierstrass function ${\cal P}(x)$
is a doubly-periodic meromorphic function, which obeys the equation
\[
{\cal P}^{\prime 2} = 4({\cal P} - e_1)({\cal P} - e_2)({\cal P} - e_3)
\equiv 4{\cal P}^3 - g_2 {\cal P} -g_3 \ .
\]
Here $e_1, e_2, e_3$ such that $e_1+ e_2+ e_3=0$ are called the roots of
the Weierstrass function.
Olshanetsky and Perelomov have shown (see, for example, \cite{Olshanetsky:1983})
that the problem (1.1) is completely integrable.

\setcounter{equation}{0}
\section{Lie-algebraic analysis}

For a two-body case we need to make a slight modification in (1.1) In particular,
adding an extra constant term
\begin{equation}
\tilde L^{(2)}\equiv \De^{(2)} - \frac{n(n+1)}{2}
\bigg[{\cal P}\bigg(\frac{x_1-x_2}{2}\bigg) + A\bigg]
\end{equation}
where $A(\om,\om')$ is a real constant with the condition that
$A(\om,\infty)= \frac{1}{3} (\frac{\pi}{2\om})^2$. Here $\om,\om'$ are
the half-periods of the ${\cal P}$--function. In particular, one can choose
$A(\om,\om')=\frac{1}{3} (\frac{\pi}{2\om})^2$. Hermiticity of $L$ requires
that one of the periods be real. Let it be $\om$.
The extra term in (2.1) just changes the reference point for energies.

Hereafter we assume that the center-of-mass coordinate $X=\sum x_i$ is separated
and the remaining part of the Laplace operator is already written
in Perelomov relative coordinates \cite{Perelomov:1971}
\begin{equation}
y_i=x_i -\frac{X}{2}\ ,\qquad i=1,2 \ ,
\end{equation}
and, say, $y_1=\frac{x_1-x_2}{2}=-y_2\equiv y$, and, thus the Laplace operator
is equal to
\begin{equation}
\Delta^{(2)} = 2 \pa_{XX}^2 + \frac{1}{2} \pa_{yy}^2  \ .
\end{equation}

The operator $\tilde L^{(2)}$ after separation of the center-of-mass motion
becomes
\begin{equation}
\tilde L^{(2)}_{rel}=\{2\tilde L^{(2)}\}|_{rel} =
\pa_{yy}^2 - n(n+1)[{\cal P}(y) + A] \ ,
\end{equation}
and has an important feature of coinciding either with the Hamiltonian of
Calogero model (both periods tend to infinity, thus $A=0$) or to the
Sutherland model (if the second [complex] period tends to infinity while
$\om$ is kept fixed; in this case $A=\frac{3g_3}{2g_2}$). This operator
(2.4) is the Weierstrass form of the Lame operator from textbooks (see, for
example, \cite{Kamke:1959, Arscott:1964}).

In the rest of the analysis we assume that $n$ is an even number. We study pure
polynomial solutions (non-vanishing at the points, where
${\cal P}(y)=e_i,\ i=1,2,3$). These solutions are called the Lame polynomials
of the first type (see, for example, \cite{Kamke:1959}).
Implicitly, it means that we are exploring periodic-periodic solutions of the
Lame equation.

The well-known algebraic form of the Lame operator (2.4) occurs if a new variable,
$\xi={\cal P}(y)$, is introduced:
\begin{equation}
\tilde L^{(2)}_{rel}= (4 \xi^3 - g_2 \xi -g_3) \pa_{\xi\xi}^2
+ (6 \xi^2 - \frac{g_2}{2})\pa_{\xi} - n(n+1)(\xi + A) \ .
\end{equation}
The operator (2.5) can be immediately rewritten in terms of the generators
of the $sl(2)$-algebra:
\begin{equation}
J^{+}_k = \rho^2 {\pa \over \pa \rho} - k \rho\ ,
\ J^{0}_k = \rho {\pa \over \pa \rho} - {k \over 2}\ ,
\ J^{-}_k = {\pa \over \pa \rho}\ ,
\end{equation}
and then the following expression for (2.5) emerges\footnote{$\rho$ is replaced
by $\xi$} \cite{Turbiner:1989}:
\begin{equation}
\tilde L^{(2)}_{rel}= J^{+}_{\frac{n}{2}} [4J^{0}_{\frac{n}{2}}+(3n+2)]
-g_2\bigg(J^{0}_{\frac{n}{2}}+\frac{n+2}{4}\bigg)J^{-}_{\frac{n}{2}} -
g_3 J^{-}_{\frac{n}{2}}J^{-}_{\frac{n}{2}}  \ ,
\end{equation}
in the representation (2.6) of spin $\frac{n}{2}$.
It is evident that the operator (2.5) has $(\frac{n}{2}+1)$ polynomial
eigenfunctions.

It is worth mentioning that for even values $n$ in (2.4) there also exist
three different kinds of eigenfunctions more, those proportional
to  $\eta_{ij}=(\xi - e_i)^{1/2} (\xi - e_j)^{1/2}, i \neq j=1,2,3$.
One can show that making the gauge rotation of (2.5):
$\eta_{ij}^{-1} \tilde L^{(2)}_{rel} \eta_{ij}$ we again get the operator
(2.5) with different coefficients in the part of it
which is linear function in derivative. In particular, the only new term
which occurs is $\xi\pa_{\xi}$. This operator
can be also rewritten in terms of the generators of the $sl(2)$-algebra
(2.6) but with spin $(\frac{n}{2}-1)$ \cite{Turbiner:1989,Turbiner:1994}.
This proves that the operator (2.5) contains $\frac{n}{2}$ polynomial
eigenfunctions. They are nothing but Lame polynomials of the third type.

Now we pose the question of what are the proper coordinates in these
the {\it algebraic} and also {\it Lie-algebraic} forms of the rational,
trigonometric and elliptic models occur, if corresponding limits are taken?
After some analysis we arrive at the following expression
\begin{equation}
\rho= -\frac{1}{({\cal P}(\frac{x_1-x_2}{2}) + A)}\equiv -
\frac{1}{({\cal P}(y) + A)}\ ,
\end{equation}
where, for the sake of convenience, we have put
\[
A\ \equiv \ {\al^2 \over 12} \ .
\]
In trigonometric limit $\al=\frac{\pi}{4\om}$. The variable $\rho$
obeys the following equation
\begin{equation}
(\rho')^2 = -4\rho - \al^2 \rho^2 - (12 A^2 - g_2) \rho^3 -
(4 A^3 - g_2A + g_3)\rho^4   \ .
\end{equation}
The trigonometric limit turns out to correspond to the disappearance of the
$\rho^3$ and $\rho^4$ terms, while in the rational limit the
$\rho^2$ term also vanishes. Then, in these limits, (2.4) becomes
\[
\tilde L^{(2)}_{rel}\
=\ \pa_{yy}^2 - \frac{n(n+1)\al^2}{4\sin^2(\frac{\al y}{2})} \ ,
\]
or
\[
=\ \pa_{yy}^2 - \frac{n(n+1)}{y^2}\ ,
\]
correspondingly. It is worth mentioning the `embedding chain' rule of degeneration:
\begin{equation}
\rho(y) \rightarrow {2(\cos\al y -1)\over \al^2} =
- \frac{4}{\al^2}\sin^2{\al y \over 2} \rightarrow - y^2   \ .
\end{equation}
Now we calculate the gauge rotated $\tilde L^{(2)}_{rel}$ operator in the
relative coordinate $\rho(y)$
\[
h\ =\ {\rho}^{-\frac{n}{2}}\{\pa_{yy}^2 - n(n+1)[{\cal P}(y)+A] \}
{\rho}^{\frac{n}{2}}=
\]
\[
=-\{4\rho +12 A\rho^2 +(12A^2-g_2)\rho^3+(4A^3-g_2A+g_3)\rho^4\}\pa^2_{\rho}
\]
\[
+\{2(2n-1)+12(n-1)A\rho+(n-\frac{3}{2})(12A^2-g_2)\rho^2+(n-2)(4A^3-g_2A+g_3)
\rho^3\}\pa_{\rho}
\]
\begin{equation}
-3An^2-\frac{n(n-1)}{4}(12A^2-g_2)\rho -\frac{n(n-2)}{4}(4A^3-g_2A+g_3)\rho^2
\end{equation}
This expression also can be represented in terms of $sl_2$--generators
(2.6) of spin $\frac{n}{2}$,
\[
h\ =\ -(4A^3-g_2A+g_3) J^+_{\frac{n}{2}}J^+_{\frac{n}{2}}
- (12A^2-g_2) J^+_{\frac{n}{2}}\{J^0_{\frac{n}{2}}-\frac{n(n-2)}{4}\}
\]
\begin{equation}
-12A J^0_{\frac{n}{2}} (J^0_{\frac{n}{2}}-\frac{n}{2})
-4(J^0_{\frac{n}{2}}-\frac{3n-2}{4}) J^-_{\frac{n}{2}} + \frac{3An^2}{4}
\end{equation}

It is worth noting that the gauge factor ${\rho}^{\frac{n}{2}}$ in the rational
and trigonometric limits degenerates to the Vandermonde determinant factor
or its trigonometric generalization respectively, which we also gauged away
in the Calogero and Sutherland models \cite{Ruhl:1995}.

The trigonometric limit corresponds to zeroing of two  expressions:
$12 A^2 - g_2=0$ and $4A^3 - g_2 A + g_3=0$, which leads to the vanishing
of the terms containing the positive-root $J^+$-generator. The spectrum
(2.11) is
\[
\ep_k=-3A(2k-n)^2,\ k=0,1,\ldots,\frac{n}{2}\ .
\]
Following a general criterion we conclude that the operator (2.11) preserves
the infinite flag of spaces of polynomials and the corresponding problem is
exactly solvable \cite{Turbiner:1994}. The rational limit occurs if
$A = g_2 = g_3 = 0$. Since in the rational limit the harmonic oscillator
interaction is omitted, there are no polynomial eigenfunctions. It is known
that the eigenvalues corresponding to the polynomial eigenfunctions and
the eigenfunctions themselves of (2.5), (2.11) are the branches of an
$(\frac{n}{2}+1)$-sheeted Riemann surface in the variable $g_2(g_3)$
characterized by the square-root branch points only \cite{Turbiner:1989}.
In the trigonometric limit the residues of these branch points vanish
and this surface splits off into separate sheets.

There is an intrinsic hierarchy of the algebraic form (2.11) of the
Hamiltonian (2.4): elliptic (grading +2), trigonometric (grading 0),
rational (grading -1). The elliptic case of the grading $(+1)$
\footnote{$4A^3-g_2A+g_3=0$ and hence the rhs in (2.9) is a cubic polynomial
in $\rho$} has no meaning in our study, since there is no solution of (2.9)
satisfying the `embedding chain' rule (2.10). In principle, the variable (2.8)
can be modified by allowing higher order terms in (10) and insisting on
fulfillment of (2.10). It would then become a hyper-elliptic variable.

The above analysis is performed for periodic-periodic solutions of the Lame
equation, for the case of an even number of zones, $n=2k, k=1,2,3\ldots$.
A similar study can be carried out for other types of solutions (see
discussion above), gauging away the factor $\rho^{\frac{n}{2}-1}$ in (2.4)
in addition to the factor $\eta_{ij}$. Finally, we get algebraic and
Lie-algebraic forms of the Lame equation similar to (2.11)--(2.12).
Similar results can be obtained for the case of an odd number of zones.

As a certain line of possible development one can substitute the generators
of the $gl(2|k)$-superalgebra instead of those of $sl_2$
(see \cite{Brink:1997}) and explore the matrix QM models those will occur.
We will discuss this exercise elsewhere.

\setcounter{equation}{0}
\section{Translation invariant discretization}

The next problem we study is how to discretize the Lame operator
in a translation invariant way, while simultaneously preserving the property
of polynomiality of the eigenfunctions and also isospectrality. Again we
restrict our analysis to the case of the periodic-periodic solutions and
an even number of zones, $n=2k, k=1,2,3\ldots$, respectively.

Let us introduce a translationally-covariant finite-difference operator
\begin{equation}
\label{e3.1}
{\cal D}_{+} f(x)  = \frac{(e^{\delta \pa_x} -1)}{\delta} f(x)\ =
\ \frac{f(x+\delta) - f(x)}{\delta}  \ ,
\end{equation}
where $\delta$ is the parameter. It is not difficult to show that the
canonically conjugate operator to ${\cal D}_+ f(x)$ is of the form
\cite{Smirnov:1995}
\begin{equation}
\label{e3.2}
x(1-\delta{\cal D}_-) f(x) = x e^{-\delta \pa_x} f(x)\ =
\ x f(x-\delta) \ ,
\end{equation}
where ${\cal D}_+\rightarrow {\cal D}_-$ if $\delta \rightarrow -\delta$.
Thus the operators (3.1) and (3.2) span the Heisenberg algebra, $[a,b]=1$.
One can easily show that using the operators (3.1)--(3.2)
we can construct a representation of the $sl_2$ algebra in terms of
finite-difference operators,
\[
J^+_k= x({x \over \delta}-1) e^{ - \delta \pa_x} (1 - k -
e^{ - \delta \pa_x})\ ,
\]
\begin{equation}
\label{e3.3}
J^0_k= {x \over \delta} (1 - e^{ - \delta \pa_x})-\frac{n}{2}\ ,
\ J^-= {1 \over \delta} ( e^{\delta \pa_x}-1)\ ,
\end{equation}
or, equivalently,
\[
J^+_n= x(1-\frac{x}{\delta})(\delta^2 {\cal D}_-{\cal D}_- -
(n+1)\delta{\cal D}_- + k)\ ,
\]
\begin{equation}
\label{e3.4}
J^0_k= x {\cal D}_- - \frac{k}{2}\ ,
\ J^-= \ {\cal D}_+ .
\end{equation}
In the limit $\delta \rightarrow 0$, the representation
(3.3)--(3.4) coincides with (2.6). The finite-dimensional
representation space for (2.6) and for (3.3)--(3.4)
occurring for the integer values of $k$ is the same.
Again that is a linear space of polynomials of degree not higher than $k$.

{\bf 1.}\ It is quite obvious that any operator written in terms of the
generators of the algebra $sl_2$ has the same eigenvalues for both
representations (2.6) and (3.3)--(3.4) (for the proof and a
discussion see \cite{Smirnov:1995}). Let us take the operator (2.5) and
construct the isospectral finite-difference operator. In order to do it
we take the representation (2.7) of this operator and substitute the
$sl_2$ generators in the form (3.3) or (3.4). Finally,
we arrive at the following finite-difference operator,
\[
{4 \xi^{(3)} \over \de^2} e^{-3 \de\pa_{\xi}} \ -2 \frac{\xi^{(2)}}{\de}
\bigg( 4\frac{\xi}{\de}-5\bigg)e^{-2 \de \pa_{\xi}} \ +
\bigg\{4\frac{\xi^{(3)}}{\de^2} + 6\frac{\xi^{(2)}}{\de} -
\bigg[\frac{g_2}{\de^2} n(n+1)\bigg]\xi\bigg\} e^{ -\de \pa_{\xi}}
\]
\[
+\ 2\frac{g_2}{\de^2}\xi + \frac{g_2}{2\de}-\frac{g_3}{\de^2}-n(n+1)A
\]
\begin{equation}
\label{e3.5}
-\ \bigg(\frac{g_2}{\de^2}\xi+\frac{g_2}{2\de}-2\frac{g_3}{\de^2}\bigg)
e^{\de\pa_{\xi}}\ -\ \frac{g_3}{\de^2} e^{2\de\pa_{\xi}} \ .
\end{equation}
The spectral problem corresponding to the operator (3.5) can be written as
the 6-point finite-difference equation
\[
{4 \xi^{(3)} \over \de^2} f(\xi-3\de) \ -2 \frac{\xi^{(2)}}{\de}
\bigg( 4\frac{\xi}{\de}-5\bigg)f(\xi-2\de) \ +
\]
\[
\bigg\{4\frac{\xi^{(3)}}{\de^2} + 6\frac{\xi^{(2)}}{\de} - [\frac{g_2}{\de^2}
n(n+1)]\xi\bigg\} f(\xi-\de)\
+\ \bigg[2\frac{g_2}{\de^2}\xi + \frac{g_2}{2\de}-\frac{g_3}{\de^2}-n(n+1)A
\bigg]f(\xi)
\]
\begin{equation}
\label{e3.6}
-\ \bigg(\frac{g_2}{\de^2}\xi+\frac{g_2}{2\de}-2\frac{g_3}{\de^2}\bigg)
f(\xi+\de)\ -\ \frac{g_3}{\de^2} f(\xi+2\de) \ =\ \la f(\xi) \ ,
\end{equation}
where $\la$ is the spectral parameter and
$\xi^{(n+1)}=\xi(\xi - \de) (\xi - 2\de) \ldots (\xi - n\de)$ is the
so called `quasi-monomial'.

Polynomial solutions of the equation (3.6) are associated with the Lame
polynomials of the first type. They can be constructed as
\begin{equation}
\label{e3.7}
\tilde P_{\frac{n}{2}}(\xi) = \sum_{i=0}^{\frac{n}{2}} \ell_i \xi^{(i)}\ ,
\end{equation}
where $\ell_i$ are the coefficients the Lame polynomial of the first
type. We call these polynomials `the 1-associated finite-difference
Lame polynomials of the first type'. One show that there exist slightly
modified 6-point finite-difference equations of the same functional form
as (3.6) having polynomial solutions of the form (3.7) associated with
the Lame polynomials of the second, third and fourth types. All their eigenvalues
corresponding to the polynomial eigenfunctions have no dependence on the
parameter $\de$.

{\bf 2.} Now we proceed to study the translation invariant
discretization of the operator (2.11). We substitute the $sl_2$-generators
of the form (3.3)--(3.4) into (2.12) and arrive at
\[
-(4A^3-g_2A+g_3){\rho^{(4)} \over \de^2} e^{-4 \de\pa_\rho}
\]
\[
+\ \bigg[2(4A^3-g_2A+g_3)\bigg(\frac{\rho}{\de}-\frac{n}{2}-2\bigg)-
(12A^2-g_2)\bigg]\frac{\rho^{(3)}}{\de} e^{-3 \de \pa_\rho}
\]
\[
-\ \bigg\{(4A^3-g_2A+g_3)\de^2
\bigg(\frac{\rho}{\de}-\frac{n}{2}-1\bigg)\bigg(\frac{\rho}{\de}-
\frac{n}{2}-2\bigg)
\]
\[
-\ (12A^2-g_2)\de\bigg(2 \frac{\rho}{\de}-n -\frac{5}{2}\bigg)-12A \bigg\}
\frac{\rho^{(2)}}{\de^2}e^{-2 \de \pa_\rho}
\]
\[
-\ \bigg\{(12A^2-g_2)\de
\bigg(\frac{\rho}{\de}-\frac{n}{2}-\frac{1}{2}\bigg)
\bigg(\frac{\rho}{\de}-\frac{n}{2}-1\bigg)
-24A\bigg(\frac{\rho}{\de}-\frac{n}{2}-\frac{1}{2}\bigg) +\frac{4}{\de}\bigg\}
\frac{\rho}{\de} e^{ -\de \pa_\rho}
\]
\[
-\ \bigg[12A\bigg(\frac{\rho^2}{\de^2}-n\frac{\rho}{\de}+\frac{n^2}{4}\bigg)-
8\frac{\rho}{\de^2}+\frac{2(2n-1)}{\de}\bigg]
\]
\begin{equation}
\label{e3.8}
-\frac{4}{\de}\bigg(\frac{\rho}{\de} -n+\frac{1}{2}\bigg)e^{\de\pa_\rho}\ .
\end{equation}
The spectral problem corresponding to the operator (3.8) can be written as
the 6-point finite-difference equation
\[
-(4A^3-g_2A+g_3){\rho^{(4)} \over \de^2} f(\rho-4\de)
\]
\[
+\ \bigg[2(4A^3-g_2A+g_3)\bigg(\frac{\rho}{\de}-\frac{n}{2}-2\bigg)-
(12A^2-g_2)\bigg]\frac{\rho^{(3)}}{\de} f(\rho-3\de)
\]
\[
-\ \bigg\{(4A^3-g_2A+g_3)\de^2
\bigg(\frac{\rho}{\de}-\frac{n}{2}-1\bigg)\bigg(\frac{\rho}{\de}-\frac{n}{2}-2\bigg)
\]
\[
-\ (12A^2-g_2)\de\bigg(2 \frac{\rho}{\de}-n -\frac{5}{2}\bigg)-12A \bigg\}
\frac{\rho^{(2)}}{\de^2}f(\rho-2\de)
\]
\[
-\ \bigg\{(12A^2-g_2)\de
\bigg(\frac{\rho}{\de}-\frac{n}{2}-\frac{1}{2}\bigg)
\bigg(\frac{\rho}{\de}-\frac{n}{2}-1\bigg)
-24A\bigg(\frac{\rho}{\de}-\frac{n}{2}-\frac{1}{2}\bigg) +\frac{4}{\de}\bigg\}
\frac{\rho}{\de} f(\rho-\de)
\]
\[
-\ \bigg[12A\bigg(\frac{\rho^2}{\de^2}-n\frac{\rho}{\de}+\frac{n^2}{4}\bigg)-
8\frac{\rho}{\de^2}+\frac{2(2n-1)}{\de}\bigg] f(\rho)
\]
\begin{equation}
\label{e3.9}
-\frac{4}{\de}\bigg(\frac{\rho}{\de} -n+\frac{1}{2}\bigg)f(x+\de)\ =\ \la f(\rho)\ .
\end{equation}

Polynomial solutions of the equation (3.9) are associated with the Lame
polynomials of the first type. They can be constructed as
\begin{equation}
\label{e3.10}
\tilde P_{\frac{n}{2}}(\rho)\ =\ \sum_{i=0}^{\frac{n}{2}} \ell_{\frac{n}{2}-i}
\rho^{(i)}\ ,
\end{equation}
(cf.(3.7)), where $\ell_i$ are the coefficients of the Lame polynomial of the first
type. We call these polynomials `the 2-associated finite-difference
Lame polynomials of the first type'. One can show that there exist slightly
modified 6-point finite-difference equations of the same functional form
as (3.9) having polynomial solutions of the form (3.10) associated with
the Lame polynomials of the second, third and fourth types. All their eigenvalues
corresponding to the polynomial eigenfunctions have no dependence on the
parameter $\de$. In the trigonometric limit the coefficients in front of
the $f(\rho-4\de), f(\rho-3\de)$ terms disappear and (3.9) becomes the 4-point
finite-difference equation.

\setcounter{equation}{0}
\section{Dilatation invariant discretization}

Now we proceed to study a dilatationally-invariant discretization of the
Lame operator (2.4). The prescription of the discretization we are going to
use is a preservation of the polynomiality of the eigenfunctions under
discretization.

Let us introduce a dilatationally-invariant, finite-difference operator.
One of the simplest operator possessing such a property is known in the literature
as the Jackson symbol (see e.g. \cite{Exton:1983, Gasper:1990})
\begin{equation}
\label{e4.1}
D f(x) = {{f(x) - f(qx)} \over {(1 - q) x}} \ ,
\end{equation}
where $q \in C$. The Leibnitz rule for the
operator $D$ is
\[
D f(x) g(x)= (D f(x)) g(x)+ f(qx) Dg(x) \ .
\]
One can construct the algebra of the first-order finite-difference operators
$D$ possessing finite-dimensional irreps for generic $q$
\cite{Ogievetsky:1991},
\label{e4.2}
\[ \tilde  J^+_n = x^2 D - \{ n \} x \]
\begin{equation}
\tilde  J^0_n = \  x D - \hat{n}
\end{equation}
\[ \tilde  J^-_n = \ D , \]
where $\{n\} = {{1 - q^n}\over {1 - q}}$  is the so called $q$-number and
$\hat n \equiv {\{n\}\{n+1\}\over \{2n+2\}}$. If $n$ is non-negative
integer, the operators (4.2) possess a common finite-dimensional invariant
subspace realized by the polynomials of degree not higher than $n$,
similar to what happens to the algebra (2.6). The operators (4.2) span
the algebra $sl_{2q}$ (see the discussion in \cite{Turbiner:1994}).
In the limit $q \rightarrow 1$ the generators (4.2) become (2.6) and the
$sl_{2q}$ algebra reduces to the standard $sl_2$ algebra.

{\bf 1.}\ In \cite{Turbiner:1992} (see also \cite{Turbiner:1994})
a theorem was proven, which showed that a linear differential
(finite-difference) operator has a certain number of polynomial eigenfunctions
if and only if it admits a representation in terms of the generators (2.6)
of the $sl_2$-algebra (the generators (4.2) $sl_{2q}$-algebra).
In particular, the theorem implies that any deformation of the positive-grading
part of the operator (2.5):
\begin{equation}
\tilde L^{(2)}_{+}= 4 \xi^3\pa_{\xi\xi}^2
+ 6 \xi^2\pa_{\xi} - n(n+1)\xi\ ,
\end{equation}
must map a polynomial of degree $n/2$ onto itself
\footnote{There are no restrictions on the deformation of the numerical
coefficients in the remaining part of the operator (2.6) -- they can be
deformed in any way we like}.
A minimal and perhaps natural deformation of the operator (4.3), which occur
after replacing $\pa_{\xi}\rightarrow D_{\xi}$ is given by
\begin{equation}
\tilde L^{(2)}_{+,def}= 4\xi^3 D_{\xi\xi}^2 + 6 \xi^2 D_{\xi} - 4
\bigg\{\frac{n}{2}\bigg\}\bigg[\bigg\{\frac{n}{2}-1\bigg\}+\frac{3}{2}\bigg]
\xi\ ,
\end{equation}
which can be immediately rewritten as a linear combination of $\tilde
J^+_{\frac{n}{2}} \tilde J^0_{\frac{n}{2}}$ and $\tilde J^+_{\frac{n}{2}}$
terms.  In general this result can be treated in such a way that the
$q$-deformation of the Lame equation (2.4) is reduced to the
$q$-deformation of its coupling constant
\begin{equation}
n(n+1)\rightarrow 4
\bigg\{\frac{n}{2}\bigg\}\bigg[\bigg\{\frac{n}{2}-1\bigg\}+\frac{3}{2}\bigg]\ .
\end{equation}
Similar deformations can be carried out for other algebraic forms of the
Lame operator associated with the Lame polynomials of the second, third and
fourth types.

{\bf 2.}\  In a similar way one can carry out a deformation of another
algebraic form of the Lame operator (2.11). Again a deformation of only the
positive-grading part of (2.11)
\[
h_+\ =\ -[(12A^2-g_2)\rho^3+(4A^3-g_2A+g_3)\rho^4]\pa^2_{\rho}
\]
\[
+[(n-\frac{3}{2})(12A^2-g_2)\rho^2+(n-2)(4A^3-g_2A+g_3)\rho^3]\pa_{\rho}
\]
\begin{equation}
-\frac{n(n-1)}{4}(12A^2-g_2)\rho -\frac{n(n-2)}{4}(4A^3-g_2A+g_3)\rho^2\ ,
\end{equation}
is essential for a preservation of the polynomiality of the eigenfunctions.
Under these requirement the simplest and perhaps the most natural deformation
is given by the following expression,
\[
h_{+,def}\ =\ -[(12A^2-g_2)\rho^3+(4A^3-g_2A+g_3)\rho^4]D^2_{\rho}
\]
\[
+[\bigg[\bigg\{\frac{n}{2}-1\bigg\}+\bigg\{\frac{n}{2}-\frac{1}{2}\bigg\}\bigg]
(12A^2-g_2)\rho^2+
2\bigg\{\frac{n}{2}-1\bigg\}(4A^3-g_2A+g_3)\rho^3]D_{\rho}
\]
\begin{equation}
-\bigg\{\frac{n}{2}\bigg\}\bigg\{\frac{n}{2}-\frac{1}{2}\bigg\}(12A^2-g_2)\rho-
\bigg\{\frac{n}{2}\bigg\}\bigg\{\frac{n}{2}-1\bigg\}(4A^3-g_2A+g_3)
\rho^2\ .
\end{equation}
This expression can be immediately rewritten as a linear combination of
$\tilde J^+_{\frac{n}{2}}\tilde J^+_{\frac{n}{2}}, \tilde J^+_{\frac{n}{2}}
\tilde J^0_{\frac{n}{2}}$ and $\tilde J^+_{\frac{n}{2}}$ terms.
Similar deformations can be carried out for other algebraic forms of the
Lame operator associated with the Lame polynomials of the second, third and
fourth types.

In conclusion one should mention that another dilatation-invariant
deformation of the Lame equation based on a different form of the
dilatation-invariant finite-difference operator was studied in
\cite{Wiegmann:1995}. It is worth emphasizing that this study was done
in relation to the problem of Bloch electrons in a magnetic field
(Azbel-Hofstadter problem).

\newpage
\def\href#1#2{#2}

\begingroup\raggedright\endgroup
\end{document}